\documentstyle[preprint,aps]{revtex}      
\begin{document} 

\renewcommand{\thefootnote}{\alph{footnote}}
\begin{center}
{\Large\bf The $\eta$-Nucleon Scattering Length and Effective Range\\[1ex]}

A.M.~Green\footnotemark[1]$^,$\footnotemark[3],
S.~Wycech\footnotemark[2]$^,$\footnotemark[4]\\

$\,^a$Department of Physics, University of Helsinki and Helsinki
Institute of Physics,\\ P.O. Box 9, FIN--00014, Finland\\
$\,^b$Soltan Institute for Nuclear Studies,
Warsaw, Poland
\end{center}
\setcounter{footnote}{3}
\footnotetext{email: {\tt green@phcu.helsinki.fi}}
\setcounter{footnote}{4}
\footnotetext{email: {\tt wycech@fuw.edu.pl}} 
 \setcounter{footnote}{0}
\renewcommand{\thefootnote}{\arabic{footnote}}
 
\begin{abstract}  
The coupled  $\eta N$ , $\pi N$ system is described by a K-matrix  
method. The parameters in this model are adjusted to get an optimal
fit to $\pi  N\rightarrow \pi  N$, $\pi  N\rightarrow \eta N$ and
$\gamma N\rightarrow \eta N$ data in an energy range of about 100MeV each
side of the $\eta$ threshold. \\
In the notation $T^{-1}+iq_{\eta}=1/a+\frac{r_0}{2} q^2_{\eta}+s q^4_{\eta}$,
$q_{\eta}$ being the momentum in the $\eta N$ center-of-mass,
the resulting effective range parameters 
for $\eta N$ scattering are found to be\\
$a$(fm) = 0.75(4)--i0.27(3), $r_0$(fm) = --1.50(13)--i0.24(4)\\
  and $s$(fm$^3$) = --0.10(2)--i0.01(1) 
\vskip 0.1cm


 PACS number(s): 13.75.-n, 25.80.-e, 25.40.V
\end{abstract}

\setcounter{equation}{0} 

The pion-nucleon and pion-nucleus interactions have been much studied,
both theoretically and experimentally, for many years. However, the 
corresponding interactions of the eta meson, mainly because of the lack of 
$\eta$-beams, have -- by comparison -- been neglected.
The main interest in $\eta$'s
has been the possibility of $\eta$-nuclear quasi-bound states. Such states were 
first  predicted by Haider and Liu  \cite{hai}  and Li et al. \cite{li},
when it was realised that the $\eta$-nucleon interaction was attractive.  
Calculations by Ueda indicated  \cite{ued} that this may happen already 
in the $\eta$-deuteron system. If these states exist, then one may expect 
them to be narrow in  few-nucleon systems, and so be easier to detect 
there. The first verification of this hypothesis was made by 
Wilkin \cite{wil}, who has suggested that an indirect effect 
of such a state is seen in the rapid slope of the  $ pd \rightarrow \  ^{3}He \eta$
amplitude detected just above the $\eta$ production threshold \cite{gar}. 
Also an indication of  strong three-body $\eta pp$ correlations follows
from the measurement of the  $ pp \rightarrow pp \eta$ cross 
sections in the threshold region \cite{cal}.

It has been shown in Ref. \cite{gre96} that the strengths of  $ \eta d $ 
interactions, in particular  the magnitude of the scattering length in this
system and the position of quasi-bound or virtual state are very sensitive 
to the value of the $\eta N $ scattering length. Moreover, the behaviour 
of the  $\eta N $ scattering matrix off the  energy-shell was found to be 
important.
In order that these $\eta$-nucleus studies can be put on a firmer foundation,
it is, therefore, necessary that a better parametrization of the basic $\eta$-nucleon
interaction be available. With this in mind, a three channel analysis of the  
eta-nucleon($\eta N$), pion-nucleon($\pi N$) and  the
two-pion-nucleon($\pi\pi N$) three-body system is carried
out. This is done in terms of a $K$-matrix based on  pion-nucleon 
amplitudes and  eta-production  cross sections -- the actual data being
the $\pi N$ amplitudes of Arndt et al. \cite{Arndt}, the 
$\pi N\rightarrow\eta N$ cross sections reviewed by Nefkens \cite{Nefkens}
and the $\gamma p\rightarrow\eta p$ data of Krusche et al. \cite{Krusche}.
In Ref.\cite{Schoch} it is shown that 
the photoproduction cross section runs essentially parallel to 
the electroproduction cross section in the  region some 100 MeV above 
threshold. Therefore, including electroproduction data into the following
analysis is not expected to lead to a different conclusion.

This analysis is now carried out and,  we believe,  it improves  analyses 
 done directly in terms of resonant $T$ matrices.  The main motivation for this
 study is to extract the eta-nucleon scattering length {\em and} effective 
 range and to determine in these quantities the uncertainties   allowed 
 by the existing data. The next check and refinement is expected to  
 follow from the few-body $\eta$ physics.

For  s-wave scattering in a system consisting of the two channels 
$\pi N$ and $\eta N$ -- here denoted simply by the indices $\pi$ and $\eta$ --
the $K$-matrix, which is essentially a generalisation of the scattering length,
 and the $T$-matrix that follows from it, can be written as

\begin{equation}
\label{KT}
  \hat{K} = \left( \begin{array}{ll}
 K_{\pi \pi} & K_{\eta \pi} \\
 K_{\pi \eta} & K_{\eta \eta} \end{array} \right)
 \ \ \ {\rm and} \ \
  T = \left( \begin{array}{ll}
 \frac{A_{\pi \pi}}{1-iq_{\pi}A_{\pi \pi}}&  \frac{A_{\eta \pi}}{1-iq_{\eta}A_{\eta \eta}}\\
  \frac{A_{\pi \eta}}{1-iq_{\eta}A_{\eta \eta}}&  \frac{A_{\eta \eta}}{1-iq_{\eta}A_{\eta \eta}}\end{array} \right),
\end{equation}
where $q_{\pi,\eta}$ are the center-of-mass momenta of the two mesons in the two 
channels $\pi,\eta$.
The channel scattering lengths $A_{ij}$ are expressed in terms of the 
$K$-matrix elements, via the solution of $T=K+iKqT$,
\begin{center}
$A_{\pi \pi}=K_{\pi \pi}+iK^2_{\pi \eta}q_{\eta}/(1-iq_{\eta}K_{\eta \eta})$,
 \ \ $ A_{\eta \pi}=K_{\eta \pi}/(1-iq_{\pi}K_{\pi \pi}) $
\end{center}
\begin{equation}
\label{2.2}                                 
A_{\eta \eta}=K_{\eta \eta}+iK^2_{\eta \pi}q_{\pi}/(1-iq_{\pi}K_{\pi \pi}).
\end{equation}
These equations form a basis in which to describe two channel scattering
in terms of the  parameters of the $K$-matrix. These $K$-matrices must
account for several observed features of the experimental data -- in
particular: 

a) The  $S$-wave $\pi N$ resonances $S(1535)$ and $S(1650)$. The effect of
these is
inserted as poles at $E=E_0$ and $E_1$, which are treated as free
parameters. However, their values are expected to be near 1535MeV and 1650MeV.
Differences arise, since $E_{0,1}$ are renormalised by the presence of 
two background terms $K_{\pi\eta}$ and $K_{\eta \eta}$,
which describe  other forms of the interactions and channel couplings not
included explicitly.

b) Experimentally, the $\eta$ does not appear to couple to the  $S(1650)$
resonance, and so this coupling is not included in the model.

c) There is  a small correction for inelasticities of the $S(1535)$ and $S(1650)$ 
due to couplings to the two-pion nucleon channel. This is  treated
 in an " optical potential manner", which means the 
introduction of the two-pion channel $K$ matrix and its subsequent 
elimination. This leads to a complex correction to the two channel  
$K$ matrix. For example, the  $S(1535)$ has a coupling to the three-body channel  
described by a singular $K$-matrix  
$K_{3,3}= \frac{\gamma_{3} \gamma_{3}}{E_0-E}$
and its coupling to the two-body channels  by 
$ K_{3,i}= \frac{\gamma_{3} \gamma_{i}}{E_0-E}$, where $i=\pi,\eta$. 
In the three body-channel, 
there is a relative momentum  equivalent to the above $ q_{i}$. This
is a three-body phase space element $ q_{3}$. It may be included together
with the coupling parameter $ \gamma_{3}$ into a small contribution to the
width of the $S(1535)$ width  
$\Gamma_{\pi ,\pi}/2 =   \gamma_{3} q_{3}\gamma_{3} $.
Only this combination of the two-pion parameters enters the correction  
to the $K$ matrix of the two channel problem. In principle, it should be  
proportional to the three body phase space, and this energy dependence 
has been accounted for. Now, the correction to the basic two channel 
$K$ matrix, which stems from the three body channel is readily 
obtained to be
\begin{equation}
\label{k3}
 \delta K_{i,j}^0=
 i \frac{K_{i,3} q_{3} K_{3,j}} { 1- iq_{3} K_{3,3}}. 
\end{equation}
A similar procedure is applied to describe the slightly higher inelasticity of 
the $S(1650)$ resonance.

These features are included in the $K$-matrices as follows.
\begin{center}
$ K_{\pi\pi}\rightarrow \frac{\gamma_{\pi}(0)}{E_0-E}+
\frac{\gamma_{\pi}(1)}{E_1-E}+i\frac{K_{\pi 3}q_3K_{3 \pi}}{1-iq_3K_{33}}$ ,
 \ \ $ K_{\pi\eta}\rightarrow K_{\pi\eta}+\frac{\sqrt{\gamma_{\pi}(0)\gamma_{\eta}}}{E_0-E}
+i\frac{K_{\pi 3}q_3K_{3 \eta}}{1-iq_3K_{33}},$
\end{center}
\begin{equation}
 K_{\eta\eta}\rightarrow K_{\eta\eta}+\frac{\gamma_{\eta}}{E_0-E}
+i\frac{K_{\eta 3}q_3K_{3 \eta}}{1-iq_3K_{33}},
\end{equation}
where \ \ \ \ \ \ \ \ \ \ 
$ K_{33}= \frac{\gamma_3(0)}{E_0-E}+\frac{\gamma_3(1)}{E_1-E}$ ,
\ \ \ \ \ $ K_{\pi 3}= \frac{\sqrt{\gamma_{\pi}(0)\gamma_{3}(0)}}{E_0-E} 
+\frac{\sqrt{\gamma_{\pi}(1)\gamma_{3}(1)}}{E_1-E},$
\begin{equation}
 K_{\eta 3}= \frac{\sqrt{\gamma_{\eta}\gamma_{3}(0)}}{E_0-E}.
\end{equation}
In the above model, there are 10 parameters that are determined by a Minuit fit to 110
pieces of data -- 23 are $\pi N$ amplitudes (real and imaginary)\cite{Arndt}, 
11 are $\pi N\rightarrow \eta N$ cross-sections[$\sigma(\pi\eta)$]
\cite{Nefkens} and  53 are $\gamma N\rightarrow \eta N$ crossections
[$\sigma(\gamma\eta)$]\cite{Krusche}. 
In practice, the actual cross section data was  used in a
reduced form, from which threshold factors have been removed -- namely:
\begin{equation}
\sigma(\pi\eta)_r=\sigma(\pi\eta)\frac{q_{\pi}}{q_{\eta}} \ \ \ \ 
{\rm and} \ \ \ \                                           
\tau(\gamma\eta)_r=\sqrt{\sigma(\gamma\eta)\frac{E_{\gamma}}{4\pi q_{\eta}}}.
\end{equation}

\noindent The values of $\tau(\gamma\eta)_r$ given in Ref. \cite{Krusche} are
used here directly, even though the mass of the $\eta$ there is 547.12MeV,
compared with the present value of 547.45MeV  in Ref.\cite{PDT}.
Such small differences are unimportant here, since the main threshold
effect is removed by considering the combination
$\sigma / q_{\eta}$.
In terms of the scattering amplitudes ($T$) of eq.(\ref{KT}), the 
corresponding model expressions are:

\[\sigma(\pi\eta)_r=4\pi\left[(Re \ T_{\pi\eta})^2+(Im \ T_{\pi\eta})^2\right]
\frac{2q_{\pi}}{3q_{\eta}} \ \ \ \ 
{\rm and} \ \ \ \                                            
\tau(\gamma\eta)_r=A(Phot) \sqrt{(Re \ T_{\eta\eta})^2+(Im \ T_{\eta\eta})^2},\]
where $A(Phot)$ is a normalisation parameter that simulates the actual
production amplitude. This parameter is assumed to be energy independent and is
treated as a free parameter in the Minuit minimization.
The resulting fit had a $\chi^2$ of 0.83/dof and the outcome is seen in Fig. 1.
Since it is not clear that the four sets of data in Fig. 1 have equal weight,
it is of interest to also look at the separate $\chi^2$/dpt --
A) 0.73 B) 0.75 C) 0.94 and D) 0.60. This shows that, indeed, good fits
are achieved in all four sets of data and that the overall $\chi^2$/dof is not
dominated by any particular set.

From Table 1 it is seen that those parameters that can be compared with
numbers in the Particle Data Tables\cite{PDT} fall into three classes:

a) $\Gamma(Total), \ \eta(br), \ \pi(br), \ \Gamma(Total,1)$ and $\pi(br,1)$ 
can be compared directly and are seen to be consistent with the experimental 
uncertainties. 
The relationship between the above $\Gamma$'s and the $\gamma$'s 
is determined by the $ T $ matrix,  which -- close to the resonance -- should be
of  a Breit-Wigner form with an energy dependent width. This relates the  
channel parameters $ \gamma $ to the total width $ \Gamma $ , with elasticities 
and the channel momenta calculated at the  resonance energy $ q(PDT)$. 
Thus, for example $\gamma_{\pi}= 0.5\pi(br)\Gamma/q(PDT)$.

b) $E_0$ and $E_1$ are the positions of the bare poles in the $K$-matrices.
As mentioned earlier, these get slightly renormalised in going from 
$K$-matrices to $T$-matrices to give the
numbers in the Particle Data Tables\cite{PDT}.

c) The seven parameters in a,b) are essentially obtained by fine-tuning
the corresponding experimental numbers -- as is seen by the close
agreement between the two. However, the remaining three parameters
$K_{\eta \eta}$, $K_{\pi \eta}$  and $A(Phot)$ are
completely free. In principle, the first two could be related to some more 
fundamental model  based on some underlying lagrangian as in Refs. \cite{sau}.
The third parameter could also be calculated, if a mechanism for $\eta$
photoproduction were used.

The values of the branching ratios $\eta(br), \ \pi(br)$ for the
$S(1535)$ resonance also give a prediction
for the two-pion ratio to be $1-\eta(br)-\pi(br)=0.038$ -- a number in line
with  experimental estimates of 0.05--0.20.

The errors on $a, \ r_0$ and $s$ were obtained by repeating the calculation
for a random selection of the nine  parameters defining the $K$-matrices of
eq.(\ref{KT}). This selection was
chosen to ensure the distribution of each parameter was a gaussian centered on
the values in Table \ref{table1} and with the same standard deviation.
Several tests were made to determine the dependence of these errors on the 
number of runs and on the size of the region each side of the gaussian maximum
over which the random points were chosen. The errors shown are for 1000 runs
using regions that were 3 standard deviations.

The negative sign of the effective range is expected, since it arises quite
naturally due to the proximity of the $S(1535)$.
For the single $\eta$ channel case dominated by the resonance, one 
would have  $ r_0 q^2_{\eta}/2 = (E_{threshold} -E )/ {\gamma_{\eta}}$.
This is  a fairly large negative effective range of about --3 fm. The 
presence of  other channels and background terms reduce it to about 
half of this value.
The shape parameter appears to be small. In fact the imaginary part is 
consistent with zero.

In Table 2 a comparison is made with earlier determinations of the scattering
length. There it is seen that the present result supports, in particular, the
estimates of Refs.\cite{wil} and \cite{aba96}. It is difficult to compare with
the other references, since they do not give any error estimates.

\noindent Fig. 2 shows that, within 30MeV of the $\eta$ threshold, the effective 
range expansion is very good. For a parametrization up to 100MeV from the 
threshold, the effect of the shape parameter($s$) plays an increasingly
important role.  Also it is seen that the effective range must be included,
if the $\eta N$ scattering is needed 10--20MeV away from the $\eta$ threshold
at 1485.7MeV. Such excursions from the threshold are needed, for example, when 
extrapolating below the threshold in $\eta$-bound state situations. In the
present case, the threshold value of the $\eta N$ amplitude (0.75+i0.27)
becomes 0.49+i0.10 fm at 1468.4MeV and 0.51+i0.51 fm at 1500.0MeV.
Such differences could be crucial in discussions concerning the existence,
or not, of $\eta$-bound states in few nucleon systems.

One of the authors (S.W.) wishes to acknowledge the hospitality of the
Research Institute for Theoretical Physics, Helsinki, where part of this 
work was carried out. The authors also thank Drs. R. Arndt and B. Krusche
for useful correspondence and Drs. J.Niskanen and M.Sainio for several
discussions.

\newpage

\begin{table}
\begin{center}
\caption{ The optimised parameters from Minuit defining the $K$-matrices
and the corresponding values from the Particle Data Tables
(PDT)\protect\cite{PDT}.}
\vspace{1cm} 
\begin{tabular}{c|ccccc} 
&$K_{\eta \eta}$&$K_{\pi \eta}$&$E_0$(MeV)&$E_1$(MeV)&$\Gamma(Total)$(MeV)\\ \hline
Minuit&0.177(33)&0.022(13)&1541.0(1.6)&1681.6(1.6)&148.2(8.1)\\ 
PDT&--&--&1535(20)&1650(30)&150(50) \\ \hline
&$\eta(br)$&$\pi(br)$&$\Gamma(Total,1)$(MeV)&$\pi(br,1)$& $A(Phot)$\\ \hline
Minuit&0.568(11)&0.394(9)&167.9(9.4)&0.735(11)&19.74(36) \\ 
PDT&0.30--0.55&0.35--0.50&145--190&0.55--0.90&-- \\ 
\end{tabular}
\label{table1}
\end{center}
\end{table}

\begin{table}
\begin{center}
\caption{ Results compared with earlier works.
The numbers in  [...] are the values of $a$ and $r_0$, when the exact scattering 
amplitudes are fitted with $s=0$. }
\vspace{0.5cm} 
\begin{tabular}{l|c} 
Reference& Scattering Length(fm) \\ \hline
Bhalerao and Liu \protect\cite{bha}&0.27+i0.22 \\
&0.28+i0.19 \\
Benhold and Tanabe \protect\cite{ben}&0.25+i0.16\\
Arima, Shimizu and Yazaki \protect\cite{ari}& 0.980+i0.37 \\
\u{S}varc, Batinic and Slaus \protect\cite{bat}&0.886+i0.274 \\ 
Wilkin \protect\cite{wil}&0.55(20)+i0.30 \\
Sauermann et al. \protect\cite{sau}& 0.51+i0.21 \\
Abaev and Nefkens \protect\cite{aba96}& 0.621(40)+i0.306(34) \\   \hline
This paper&  \\ \hline
Scattering length($a$)& 0.751(43)+i0.274(28)\\ 
	 &[0.751(43)+i0.274(28)]\\
Effective range($r_0$)&--1.496(134)--i0.237(37)\\
&[--1.497(134)--i0.237(38)]\\
Shape parameter($s$)&--0.102(15)--i0.008(10)\\ 
\end{tabular}
\label{table2}
\end{center}
\end{table}

\newpage

\begin{figure}[ht] 
\vspace{1cm} 
\caption{ The $K$-matrix fit to experimental data as a function of the
center-of-mass energy Ecm:
A) The $\pi N\rightarrow \eta N $ data of Ref.\protect\cite{Nefkens} --
the reduced cross-section in $mb$ containing  the factor $q_{\pi}/q_{\eta}$, 
B) $\tau(\gamma\eta)_r$ the reduced cross-section of 
Ref.\protect\cite{Krusche}  in units of $10^{-3}/m_{\pi^+}$ ,
C) The real part of the $\pi N$ amplitudes ($q_{\pi} \ Re \ T$)
\protect\cite{Arndt},
D) The imaginary part of the $\pi N$ amplitudes ($q_{\pi} \ Im \ T$)
\protect\cite{Arndt}.}
\includegraphics{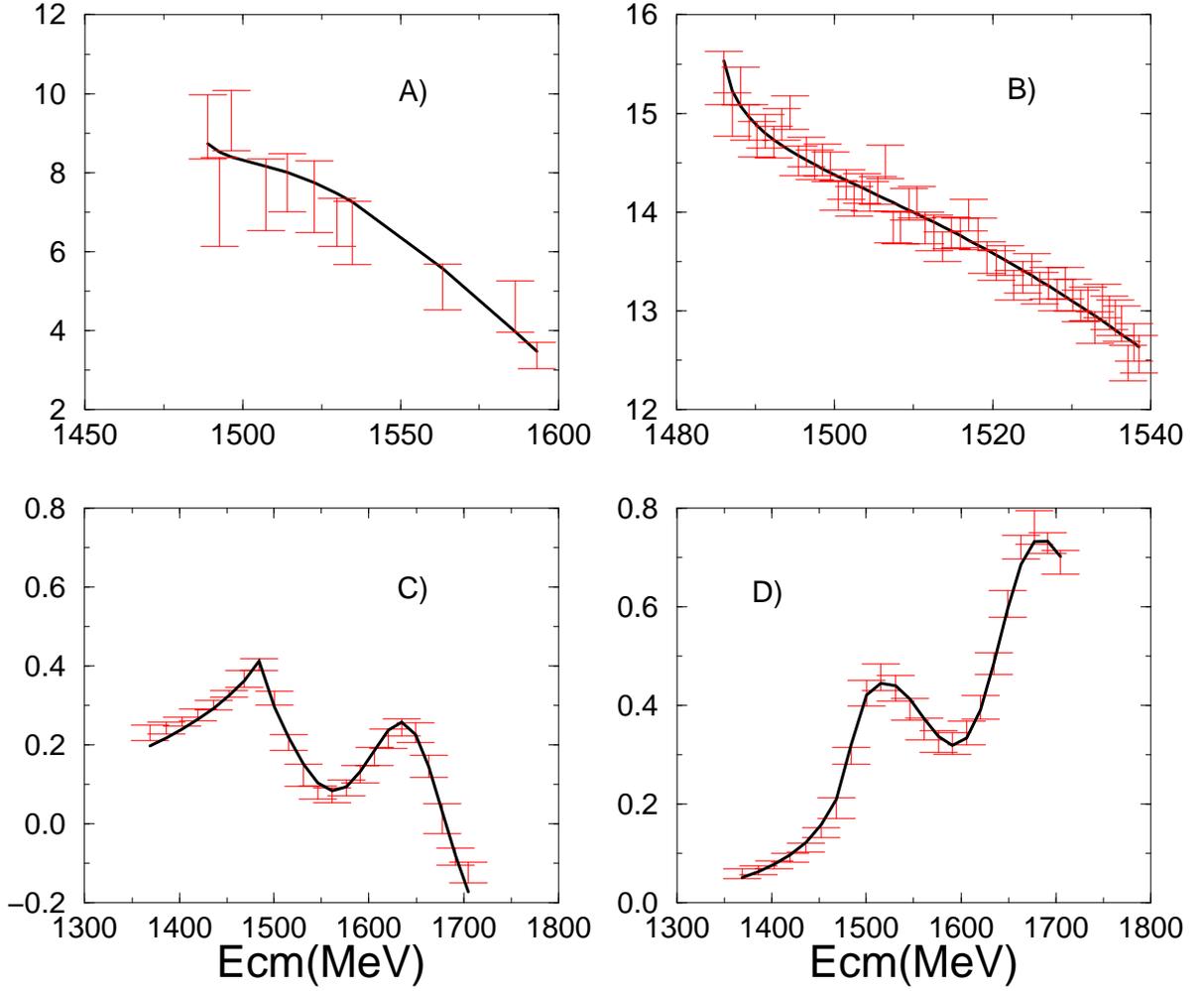}
\end{figure}


\begin{figure}[ht] 

\vspace{24cm} 

\caption{ The quality of the effective range expression versus the
exact values. The solid line shows the exact results, the dashed line
the effective range expansion with the values of $a, \ r_0, \ s$ from table 2
and the dotted line the effective range expansion with only  $a, \ r_0.$
A) shows the real parts and B) the imaginary parts. All amplitudes are in fm.}
\includegraphics{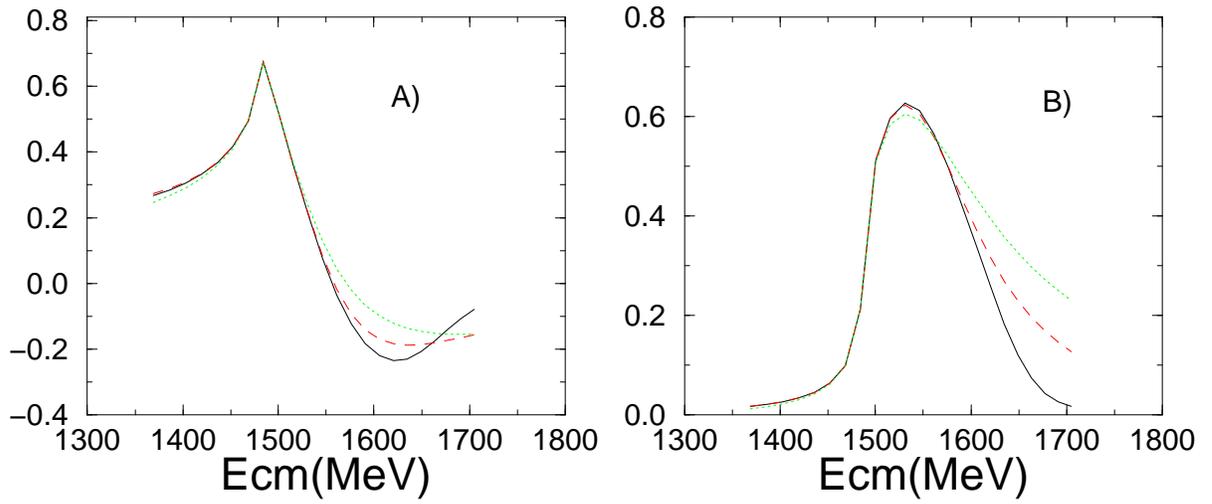}
\end{figure}

\end{document}